\documentclass[fleqn,twoside]{article}
\usepackage{espcrc2}
\usepackage{latexsym}

\usepackage{graphicx}

\newcommand{\AmS}{{\protect\the\textfont2
  A\kern-.1667em\lower.5ex\hbox{M}\kern-.125emS}}

\title{Topics of monopole dynamics in gluodynamics}

\author{T. Suzuki,\address[KNZ]{Inst. Theor. Phys., Kanazawa Univ., Kanazawa
920-1192, Japan}
        Y. Koma,\addressmark[KNZ]
        S. Ito,\addressmark[KNZ]
        E.-M. Ilgenfritz,\address[RCNP]{RCNP, Osaka Univ., Osaka 567-0047,
	Japan}
        T.W. Park,\addressmark[KNZ]
        M.I. Polikarpov,\address[ITEP]{ITEP, Moscow,117259, Russia} and 
        T. Yazawa.\address[KJC]{Kinjo College, Ishikawa 924-8511, Japan}}

\begin{document}

\begin{abstract}
Three topics of monopole dynamics in gluodynamics are presented.
1)Gauge (in)dependence of the monopole effective action
(S.Ito, T.W.Park, T.Suzuki):
Four different abelian projections are studied. MA and Laplacian abelian
 gauges show almost the same renormalization flow and renormalized trajectory.
2)The quantized dual abelian Higgs model (DAH) derived from SU(2) gluodynamics 
and its vacuum structure (Y.Koma, E.-M.Ilgenfritz,
 T.Suzuki,M.I.Polikarpov):
Monte-Carlo analysis of DAH is done. The quantum average of the dual
 field strength shows a flux tube profile similar to the color-electric field profile
measured in SU(2) gluodynamics.
3)Lattice instanton action from 3D SU(2) Georgi-Glashow model (GG)$_3$
(T.Yazawa, T.Suzuki):
(GG)$_3$ is studied on the lattice in the London limit.
We determine an effective instanton action both
in unitary and MA gauges. For some range of parameters,
we obtain almost perfect actions which look the same in both gauges 
(gauge independence) and which reproduce well the string tension.
\end{abstract}

\maketitle

\section{Gauge (in)dependence of the monopole effective action in SU(2)}
Four different abelian projections are studied.
Maximally abelian gauge (MAG) is given by 
diagonalizing the following operator
\begin{small}
\begin{eqnarray*}
X(x)=\sum_\mu\left[U_\mu(x)\sigma_3U_\mu^\dagger(x)+
U_\mu^\dagger(x-\hat{\mu})\sigma_3U_\mu(x-\hat{\mu})\right].
\end{eqnarray*}
\end{small}

Polyakov gauge and F${}_{12}$ gauge are defined
with diagonalizing the operators
\begin{eqnarray*}
X_{Pol}(x)&=&\prod_{i=1}^{N_4}U_4(x+(i-1)\hat{4}),\\
X_{F_{12}}(x)&=&
U_1(x)U_2(x+\hat{1})U_1^\dagger(x+\hat{2})U_2^\dagger(x),
\end{eqnarray*}
respectively.

The gauge fixing matrix $\Omega$  
in the Laplacian abelian gauge (LAG)~\cite{sijs} is
determined as 
\begin{eqnarray*}
\Omega^\dagger(x)\sigma_3\Omega(x)=\hat{\phi}_a(x)\sigma_a,
\end{eqnarray*}
where $\phi$  is the eigenvector
belonging to the lowest eigenvalue of the covariant Laplacian
$\Box_{xy}^{ab}$:
\begin{small}
\begin{eqnarray*}
-\Box_{xy}^{ab}=&&\hspace*{-1cm}\sum_\mu\left(
2\delta_{xy}\delta^{ab}-R_\mu^{ab}(x)\delta_{y,x+\hat{\mu}}
-R_\mu^{ba}(y)\delta_{y,x-\hat{\mu}}\right),\\
R_\mu^{ab}(x)&=&\frac{1}{2}\mbox{Tr}\left(
\sigma_aU_\mu(x)\sigma_bU_\mu^\dagger(x)\right).
\end{eqnarray*}
\end{small}

\vspace{-.5cm}

An effective monopole action $S(k)$ is determined by the
Swendsen's inverse M-C method~\cite{ss}.
We adopt 27 two-point+4- and 6-point interaction terms 
as the form of action~\cite{kato}.
\begin{eqnarray*}
S[k]=\sum_{i=1}^{27}g_iS_i[k]+g_{28}(k_\mu^2(x))^2+g_{29}(k_\mu^2(x))^3.
\end{eqnarray*}
Running of the coupling constants are determined by the effective
actions $S^{(n)}[k^{(n)}]$ fixed from configurations of blocked 
monopole currents $k^{(n)}$:
\[ S[k]\rightarrow S^{(2)}[k^{(2)}]\rightarrow\cdots
\rightarrow S^{(n)}[k^{(n)}]\rightarrow\cdots \]

\begin{figure}[htb]
\includegraphics[scale=0.35]{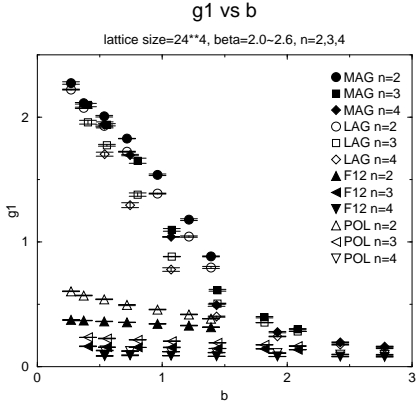}
\label{fig:g1}
\vspace*{-0.9cm}
\caption{$G1$ vs $b$ in various gauges for $n=2\sim4$ steps of blocking 
 on $24^4$.}
\vspace*{-0.5cm}
\end{figure}

Figure \ref{fig:g1} shows the most dominant self coupling versus $b$ in
all gauges adopted.
In the case of  MAG and  LAG,  the scaling behavior (a unique curve for
different $n$ ) is seen and 
both coupling constants are very close to each other. However, the
scaling behavior is not seen for small $b$ in
other two unitary gauges. We need more steps of the block-spin transformation.

\section{The quantized dual abelian Higgs model derived from SU(2)
 gluodynamics and its vacuum structure}
We have studied the quantitative relation between
Abelian projected SU(2) gluodynamics and 
the {\it quantum} U(1) dual Abelian Higgs model (DAHM). 
The lattice action is 
\begin{eqnarray*}
&&
S_{\rm DAH}[B,\chi,\chi^{*}]= 
\sum_{s}
\Biggl [
\frac{\beta_{g}}{2} \sum_{\mu < \nu} {}^{*\!}F_{s,\mu \nu}^2
\nonumber\\
&&
+\gamma \sum_{\mu} \left | \chi_{s} - e^{i B_{s,\mu}}
\chi_{s+\hat{\mu}}\right |^2
 +
\lambda \left ( |\chi_{s}|^2 - 1 \right )^2 \Biggr ],\nonumber
\end{eqnarray*}
with the dual field strength ${}^{*\!} F_{s,\mu \nu}
=
B_{s,\mu} + B_{s+\hat{\mu},\nu}
-B_{s+\hat{\nu},\mu}-  B_{s,\nu}$,  the dual gauge field $B_{s,\mu}$
and $\chi_{s}$ as a complex monopole field.
We determined the bare couplings $\beta_{g}$, $\gamma$, 
and $\lambda$ by matching the monopole action extracted from 
SU(2) lattice gauge theory in MAG with the effective 
monopole representation of DAHM. The latter is 
derived~\cite{Chernodub:1999xf} from DAHM by path
integration, 
keeping only the monopole currents $k_{\mu}(s)$ fixed.
The {\it form} of the two- and four-point couplings 
results from integration over the Higgs field's modulus. 
Identifying the monopoles in both approaches,
the {\it values} of the Coulomb, the 2- and 4-point couplings 
are 
obtained from monopole current networks (obtained by 
Abelian projection of SU(2) fields) using the extended Swendsen 
method~\cite{Shiba:1995pu}.

\par
Once the couplings in $S_{\rm DAH}[B,\chi,\chi^{*}]$ are fixed in
this way, for a given lattice spacing, 
one should be able to mimic SU(2) gluodynamics results
simulating DAHM as an effective theory:
(1) It would be interesting to recover the string tension by
calculating
dual 't Hooft loop expectation values and fitting them to an area law;
(2) One would like to reproduce the flux tube
profile 
(dual field strength, distribution of monopole current, the Higgs
profile). 
For both purposes, the external color-electric source 
is introduced as a Dirac string term $\Sigma_{s,\mu\nu}$ living 
in the minimal surface spanned by the color-electric current loop, by 
replacing $F_{s,\mu\nu} \to F_{s,\mu\nu}-2\pi \Sigma_{s,\mu\nu}$ 
in the action.

\begin{figure}[htb]
\includegraphics[height=13pc]{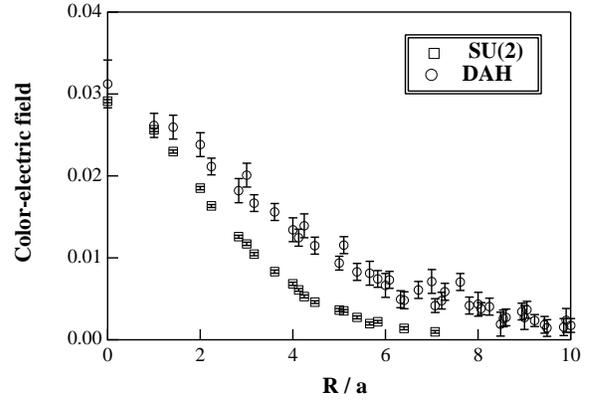}
\vspace*{-1.2cm}
\caption{The profile of the color-electric field in DAHM. 
The SU(2) result~\cite{Bali:1998gz} is also plotted for comparison.}
\label{fig:electric-SU2-DAH}
\vspace*{-0.5cm}
\end{figure}
\par
We have chosen $\beta_{\rm SU(2)} = 2.5115$ 
(adopting $a=0.086$ fm) for which the profile of the SU(2) flux tube  
has been studied in Abelian projection~\cite{Bali:1998gz}. 
For the corresponding dual lattice the bare DAHM couplings are 
$\beta_{g}=0.04$, $\gamma = 0.46$, and $\lambda = 1.17$.
With external charges included, analyzing the change $\Delta F(R,T)$
of free energy caused by rectangular 't Hooft loops,
a string tension is obtained which amounts to 86 \% of $\sigma_{\rm
SU(2)}$. 
The quantum average of the dual field strength shows a 
flux tube profile similar to the color-electric field profile
measured in Ref.~\cite{Bali:1998gz}. 
A clear signal of the monopole current encircling the flux tube
could not be obtained, due to a large vortex density found
at the actual set of coupling parameters. 
The present status suggest that monopole dynamics is really providing 
the link relating SU(2) gluodynamics to DAHM. 
For a quantitative success, however, more complicated monopole actions 
must be envisaged. Alternatively, matching could have been arranged
deeper in the infrared, on blocked lattices.

\section{Lattice instanton action from 3D SU(2) Georgi-Glashow model}
The three-dimensional Georgi-Glashow model (GG)$_3$ has
a famous \mbox{'t Hooft}-Polyakov instanton having a magnetic charge.
Polyakov showed analytically~\cite{Polyakov77} that
the string tension of 3D Georgi-Glashow model has a finite value.
He made a quasi-classical calculation  
using a dilute Coulomb gas approximation of 
\mbox{'t Hooft}-Polyakov instantons:
\begin{eqnarray*}
S= \frac{\rm const.}{g^2} \sum_{a} {q_a}^2
  +\frac{1}{4g^2} \sum_{a\neq b}
   \frac{q_a q_b \cdot 2 \pi }
       {| {\mbox{\boldmath $x$}}_a-{\mbox{\boldmath $x$}}_b |} .
\end{eqnarray*}

We study SU(2) lattice (GG)$_3$ model in the London limit
using Monte-Carlo simulation.
Abelian and instanton dominance are observed after abelian
projections in a unitary gauge ($\phi^a=\delta^{3a} $) and
MAG.
 
When we restrict ourselves to some regions of parameters $\beta$
and $\kappa$
we find that the DeGrand-Toussaint instanton~\cite{D-T} configuration realizes
the coulomb gas condition.
\begin{figure}[htb]
\vspace*{-.5cm}
\includegraphics[height=15pc]{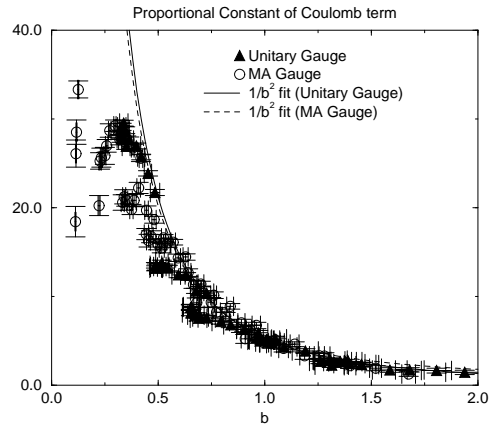}
\vspace{-25pt}
\caption{Proportional coefficient of Coulomb term.}
\label{fig:const}
\vspace*{-0.4cm}
\end{figure}

The  method is as follows. First,
the lattice instanton action is derived from instanton configuration
using inverse MC method~\cite{SW} and 
the action adopted is composed of 10 quadratic interactions:
\[
\begin{array}{rl}
 S_m =& G_0 \sum_{x} k(x)^2 + G_1 \sum_{x,\mu} k(x)k(x+\hat{\mu}) \\
     &+ G_2 \sum .... + G_9 \sum......
\end{array}
\]
This action is well fitted with the lattice Coulomb propagator($\Delta^{-1}_L$):
\[
\begin{array}{rl}
 S_m \sim& {\rm Self~term} \\
 &+ {\rm Const.}\sum_{x,x'} k(x) \Delta^{-1}_L(x-x') k(x')
~.
\end{array}
\]
Next, we perform the block-spin transformation
using the scale :
$a=\sqrt{ \sigma_L / \sigma_{\rm phys} }$ .
The proportional coefficient of Coulomb term
behave as $1/b^2$, where   $b=na$ and $n$ is
the number of block-spin transformation as seen in Fig. \ref{fig:const}.
Finally,
we obtain an almost perfect instanton action of (GG)$_3$
and it reproduces well the string tension~\cite{Yazawa}.


T.S. acknowledges financial support from JSPS
Grant-in Aid for Scientific
Research (B)  (No.11695029).


\begin{thebibliography}{99}
\bibitem{thooft}{G. 't Hooft, Nucl. Phys. B190 (1981) 455.}
\bibitem{sijs}{A.J. van der Sijs, Prog. Theor. Phys. Suppl. 131 (1998) 149.}
\bibitem{ss}{H. Shiba and T. Suzuki, Phys. Lett. B333 (1994) 461.}
\bibitem{kato}{M.N. Chernodub, {\it et al.}, Phys. Rev. D62 (2000) 094506.}
\bibitem{Chernodub:1999xf}
{M.~N. Chernodub, {\it et al.}, hep-lat/9902013.}
\bibitem{Shiba:1995pu} {H.~Shiba and T.~Suzuki, Phys. Lett. B343 
(1995) 315 ; B351 (1995) 519.}
\bibitem{Bali:1998gz}
{G.S. Bali, C.~Schlichter, and K.~Schilling,
Prog. Theor. Phys. Suppl. 131 (1998) 645.}
\bibitem{Polyakov77} {A. M. Polyakov, Nucl. Phys. B120 (1977) 429.}
\bibitem{D-T} {T.A. DeGrand and D. Toussaint, Nucl. Phys. B190 (1981) 455.}
\bibitem{SW} {R.H. Swendsen, Phys. Rev. Lett.  52 (1984) 1165.}
\bibitem{Yazawa} {T. Yazawa and T. Suzuki, JHEP 04 (2001) 026.}
\end{thebibliography}
\end{document}